\font\tenbf=cmbx10 at 12pt

\newcommand{\nonumsection}[1] {\vspace{12pt}\noindent{\tenbf #1}
        \par\vspace{5pt}}

\documentstyle[epsfig,twocolumn]{article}
\title{A Feynman graph selection tool in GRACE system}
\author{
Fukuko YUASA\(^{a} \)\footnote{E-mail:fukuko.yuasa@kek.jp},
Toshiaki KANEKO\(^{b} \),
Tadashi ISHIKAWA\(^{a} \)\\
~\\a)KEK, 1-1 OHO Tsukuba Ibaraki, Japan 305-0801\\
b)Meiji-Gakuin Univ., Kamikurata 1518, Totsuka, Yokohama,
Kanagawa Japan 244-0816
}
\date{}
\begin{document}
\maketitle

\begin{abstract}
We present a Feynman graph selection tool {\tt grcsel},
which is an interpreter written in C language.
In the framework of {\tt GRACE}, it
enables us to get a subset of Feynman graphs according to 
given conditions.
\end{abstract}
%
\section{Introduction}\label{intro}

Using an automatic Feynman graph calculation package, we can generate the
information of all Feynman graphs for given processes. 
Sometimes it is necessary to select graphs from the set of all graphs by some 
conditions. 
However, it is not so easy to select them correctly by hand
when a huge number of graphs is involved,
such as higher order corrections or SUSY processes.

A program {\tt grcsel} selects out a subset of Feynman graphs 
from the set of graphs, generated by GRACE\cite{GRACE},
according to given selection conditions.
The output information of selected graphs is written in the same
format as that of the original set. 
This enables us to generate Feynman amplitudes within GRACE
in the same procedure as for all graphs.
So we can perform cross section calculation,
gauge invariance check, event generation and so on for
the selected ones.
{\tt grcsel} helps us:
\begin{enumerate}
\item{to find decay-graphs and evaluate signal/background ratio,}
\item{to check the accuracy of approximated calculation,}
\item{to confirm precision of the  calculation,}
\item{to reduce the calculation time,}
\item{to develop kinematics routine.}
\end{enumerate}

%
\section{Overview of {\tt grcsel}}\label{overview}
{\tt grcsel}  consists of three parts: 
a steering-part defines basic functions of graph selections and
reads input files, 
an interpreter-part parses and evaluates commands, and a 
utility-part handles subsets of particles, vertices and graphs.

Once a physics process and the order of calculation are fixed, 
Feynman graphs are generated by {\tt grc} program
with specified Feynman rules described in physics model file\cite{grc}.
The information on graphs generated are
stored in a file named {\tt out.grf}
(we call the format of this file \textsl{\texttt{.grf} format}). 

\begin{figure}
\begin{center}
\rotatebox{270}{\epsfig{file=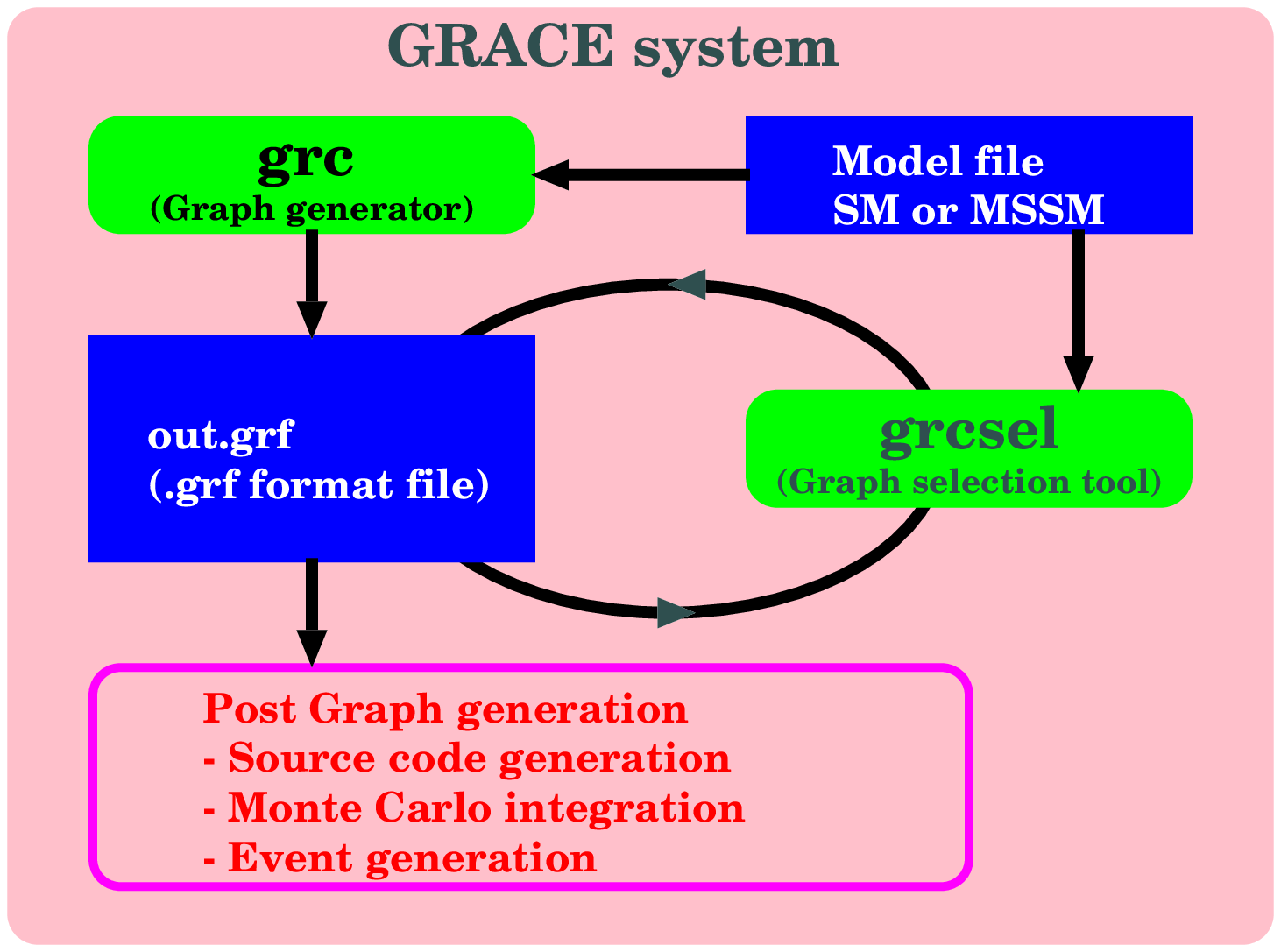,height=7cm}}
\end{center}
\caption{}
\label{fig:grcsel} 
\end{figure}

{\tt grcsel} reads the physics model file and {\tt out.grf}
and selects graphs according to a kind of propagator, 
characteristics of graph topology, a type of vertex or a graph number.
{\tt grcsel} outputs those selected graphs in the same format as
{\tt out.grf}.
Successively this output file can be used as the input to source code
generation for Monte Carlo integration or event generation.
We can also use {\tt grcsel} again reading output of
previous execution of {\tt grcsel}.
The schematic view of how {\tt grcsel} works in GRACE system is shown in 
Fig.~\ref{fig:grcsel}.

\subsection{Running {\tt grcsel}}
Graph selection starts by the program {\tt grcsel}:
\begin{center}
{\tt grcsel}
\end{center}
This program requires {\tt out.grf} file by default.
The graph selection commands are read through standard input,
which may be given interactively or by a script file.
With a script file where {\tt grcsel} commands are prepared,
we can redirect that file:
\begin{center}
{\tt grcsel < command.in}
\end{center}
To use another input {\tt .grf} format file, e.g. {\tt out1.grf}, instead of 
{\tt out.grf}, we can add the filename after {\tt grcsel} command as:
\begin{center}
{\tt grcsel out1.grf < command.in}
\end{center}

\subsection{{\tt grcsel} command}

In a script file there are a series of {\tt grcsel} commands such as
declaration of variables and basic functions to specify the selection 
conditions or operators. 
{\tt grcsel} has 14 basic functions in total.
Three of them return a subset of graphs 
in accordance with specified selection condition
and two functions output set of graphs.
They are summarized in Table~\ref{tab:func}.  
\begin{table}
\caption{Basic functions to select and output graphs.}
\label{tab:func}
\begin{center}
\begin{tabular}{|l|l|} \hline 
Function&Description\\ \hline
\texttt{cutprop}&Select graphs with\\ 
       &a specified propagator.\\ \hline
\texttt{selvlegs}&Select graphs with\\
        &a vertex consisting\\
        &specified particles.\\ \hline
\texttt{selvertex}&Select graphs with\\ 
         &a specified vertex.\\ \hline
\texttt{outgset}&Output a set of graphs.\\ \hline
\texttt{renumgset}&Renumber and output\\ 
         &a set of graphs. \\ \hline
\end{tabular}
\end{center}
\end{table}
%
 
In {\tt grcsel},
graphs, particles or vertices are  
treated as elements of a {\it set} of type
{\tt gset}, {\tt pset} or {\tt vset}, which are defined as a set of graphs, 
particles or vertices, respectively.   
Set variables have to be declared at first with their types.
Operations on sets are available and are shown 
in Table~\ref{tab:op}.
\begin{table}
\caption{{\it set} operators}
\label{tab:op}
\begin{center}
\begin{tabular}{|l|l|} \hline 
Operator& Function \\ \hline
\&    &Set intersection \\ \hline
$\mid$&Set union\\ \hline
\textasciitilde& Complement of set \\ \hline
\end{tabular}
\end{center}
\end{table}
%
%

\section{Example}\label{ex}
In the following example, graphs with $\nu_e$ propagator
connected to the initial electron and final $W^-$ are selected among graphs of 
$e^+e^- \rightarrow W^+W^-\gamma$ process. Selected graphs are 
output into a file named {\tt out1.grf}.
\begin{verbatim}
% e+ e- --> W+ W- Photon
%
% out1.grf : with neutrino propagator 
% at the vertex of initial electron and
% final W-.
%

gset gs0, gs1;

gs0 = ~[];   % all graphs

gs1 = cutprop(gs0, [``nu-e''], [0,3]);

outgset(``out1.grf'', gs1);

quit;

\end{verbatim}

\begin{figure}
\begin{center}
\epsfig{file=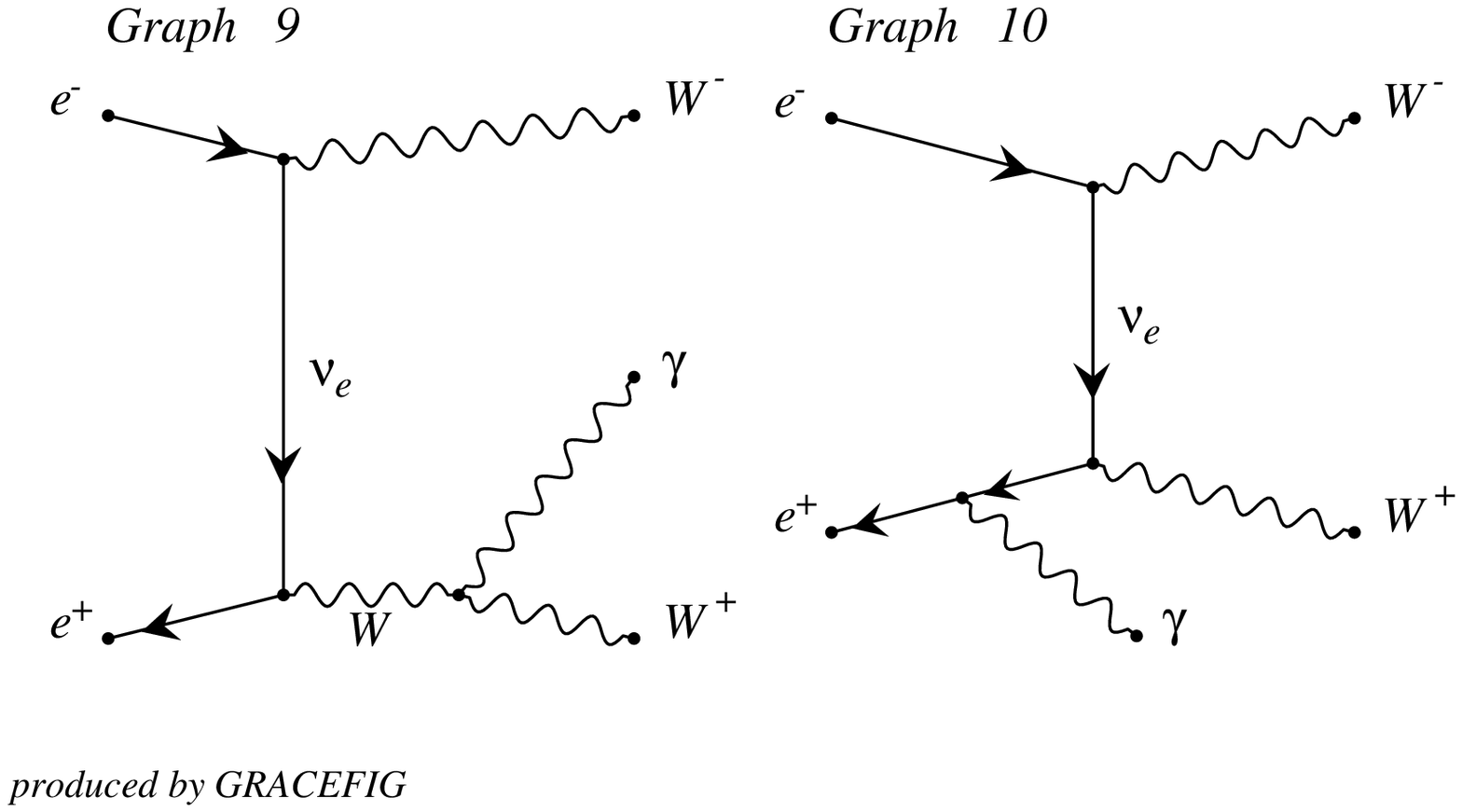,height=4.5cm}
\end{center}
\caption{}
\label{fig:wwa-select} 
\end{figure}

In Fig.~\ref{fig:wwa-select}, selected graphs are shown.

\section{Remarks}\label{remark}

{\tt grcsel} has been developed in the framework of GRACE 2.1.7.4.
It can handle tree and 1-loop graphs and it supports standard
and MSSM physics model. 
{\tt grcsel} is included in a distribution kit of GRACE 2.1.7.4.

\par
\nonumsection{Acknowledgments}
We wish to thank the members of MINAMI-TATEYA collaboration for
continuous discussions and many kinds of support.
We are also grateful to express our sincere gratitude to
Prof. Y.Shimizu for the valuable suggestions and continuous encouragements. 
Authors appreciate Prof. Y.Watase for the encouragements. 
This work was supported in part by the Grant-in Aid (No. 12680363, 10640285,
 10680366 and 11440083) of Monbu-sho, Japan. 
%

\end{document}